 \documentstyle[aps,preprint]{revtex}
\begin{document}
\draft
\title{A fresh look at 3D microwave ionization curves
of hydrogen Rydberg atoms}
\author{G. N. Rockwell, V. F. Hoffman, Th. Clausen, and R. Bl\"umel}
\address
{Department of Physics, Wesleyan University,
Middletown, CT 06459-0155}
\date{\today}
\maketitle
\begin{abstract}
Analytical arguments and numerical simulations suggest
that the shapes of
3D microwave ionization curves measured by Koch and
collaborators (see P. M.  Koch and K. A. H. van Leeuwen,
Phys. Rep. {\bf 255}, 289 (1995)) depend
only weakly on the angular momentum of the atoms
in the initial
microcanonical ensemble, but strongly on the
principal quantum number and
the magnetic quantum number.  Based on this insight,
coupled with the
computational power of a high-end 60-node Beowulf PC
cluster, we present
the first 3D quantum calculations of microwave
ionization curves in the
experimentally relevant parameter regime.
\par
\end{abstract}

\narrowtext

When Bayfield and Koch first reported their
microwave ionization
experiments \cite{BK}
the results were puzzling and appeared to be in complete
contradiction with Einstein's theory of
the photoelectric effect:
the experiments showed sharp ionization
thresholds as a function of the {\it field strength},
not the frequency, as one
might have expected.
After initial confusion
and unsuccessful attempts to
explain the results within the
framework of multiphoton theory \cite{koch},
classical molecular dynamics
calculations \cite{percival}
and classical chaos
theory \cite{meerson} emerged
as the first theories capable of providing
physical insight into the most
pertinent features of the ionization
mechanism. Thus, about a decade
after the seminal experiments by Bayfield and
Koch the reason for the surprising
behavior of the hydrogen atom in a strong
microwave field was finally
discovered: chaos. By now it is firmly
established that the hydrogen
atom in a strong microwave field is a
classically chaotic system \cite{class1,class2}.
Since nowadays quantum systems
chaotic in their classical limit are
called ``quantum chaotic''
\cite{gutzwiller}, the Bayfield and Koch
experiments were the first
quantum chaos experiments.

After an initial flurry of activity in
the field of microwave ionization of
hydrogen Rydberg atoms in the late 1970s
and in the 1980s, progress
slowed down considerably.
This is mainly due to the fact that neither the
original Bayfield and Koch experiment \cite{BK},
nor the successor
experiments by Koch and collaborators at
Stony Brook \cite{KL} are fully
quantum number resolved experiments.
Only the principal quantum number of
the Rydberg atoms is known before the
atoms are irradiated by the microwave
field \cite{KL}.
This renders these experiments effectively
three-dimensional (3D), which poses
formidable problems for classical and
quantum theory
alike. Based on new insight into the
ionization mechanism of 3D hydrogen
atoms and aided by a 60-node Beowulf cluster
computer built by the authors at
Wesleyan University \cite{WesWulf},
it has only recently
become possible to approach the 3D
microwave ionization problem of hydrogen
Rydberg atoms in the experimentally
relevant parameter regime \cite{CB}.
In this paper we report
the results of the first three-way
comparison between experiment, 3D classical
and 3D quantum theory for microwave
ionization curves of $n_0=37$ hydrogen
Rydberg atoms.

The experiments are done in the
following way \cite{KL}: protons are
generated in a plasma ion source and accelerated
to an energy of about 14
keV. In a charge exchange cell they pick up an
electron which is subsequently
excited into a high
$n_0$ Rydberg state ($n_0=24,\ldots ,90$). Following this
preparation stage the atoms fly towards
a microwave cavity. On their way
to the cavity the atoms experience stray
electric and magnetic fields.
No controlled, state-defining electric or magnetic
fields are switched
on. The stray fields are not
strong enough to change the principal quantum
number of the atoms, but
strong enough to produce a statistical mixture
of the substates of the
hydrogenic atomic shell defined by the principal
quantum number $n_0$.
This statistical mixture was demonstrated to be
consistent with a
micro-canonical ensemble \cite{KL},
i.e. all angular momentum quantum numbers
($l_0$) and magnetic
quantum numbers ($m_0$) occur
with equal probability in the initial ensemble.

Thus prepared the atoms fly
into a cylindrical microwave cavity where they
experience a total of about
500 cycles of a $9.92\,$GHz microwave field linearly
polarized along the beam axis
and operated in the TM$_{020}$ mode.
The linear polarization is an
important boon for classical and quantum calculations.
It means that $m$ is
a good quantum number. Angular momentum, however,
is strongly mixed by the
microwave field.
This can be understood intuitively on
the basis of the structure of the
angular momentum coupling matrix
$\langle lm|z|l'm\rangle$ $\sim$
$\delta_{l',l+1}+\delta_{l',l-1}$.
It is a tridiagonal matrix coupling
nearest neighbors strongly with the same weights.
Its eigenvectors are broad
states, supporting the strong
coupling argument. Thus we expect
strong $l$ mixing after only a few microwave
cycles and consequently a
microwave ionization probability that is
independent of $l_0$ to a very good
approximation.
We confirmed this analytical argument
with detailed numerical classical and
quantum calculations.
Both the analytical arguments and the
numerical simulations lead us to the
conclusion that the measured microwave ionization
threshold fields depend only weakly on
$l_0$, but strongly on $n_0$ and the
magnetic quantum number $m_0$. This is
the central physical insight that
forms the basis of the computations
reported in this paper.

In atomic units and to a good approximation
the Hamiltonian describing the experiments
by Koch {\it et al.} \cite{KL} is given by
$$
  \hat{H}=\frac{\hat{p}^2}{2} - \frac{1}{\hat{r}} +
  \epsilon \hat{z} f(t) \sin (\omega t),
\eqno(1) $$
where $\hat p$ is the electron momentum, $\hat r$
is the distance of the electron from the proton,
$\epsilon$ is the microwave
field strength, $z$ is the electron coordinate
with respect to the proton in beam direction,
$\omega$ is the microwave
frequency and
$$
f(t')=\left[1+\exp\left(-\frac{t'-92.22}{13.35}\right)
          \right]^{-1}-
  \left[1+\exp\left(-\frac{t'-409.72}{15.86}\right)\right]^{-1}
\eqno(2) $$
is an envelope function that realistically models
the switch-on and
switch-off stages of the atom upon
entering and leaving the microwave cavity
\cite{KL}. The variable $t'$ is time
measured in number of field cycles.

We perform quantum calculations in a
discrete basis of normalizable Stark states. We
choose the extremal Stark state to
represent the $l$ manifold for a given
$n_0$. The hydrogen wave function is
expanded into a set of extremal Stark
states and the coupled linear equations
resulting from the time-dependent
Schr\"odinger equation are propagated
forward in time using a fourth-order Runge-Kutta
integrator method.

Modeling the experimental situation,
ionization is implemented
via an absorbing
boundary condition
located at $n = n_c = 91$.
This is the experimental cut-off $n$ beyond which
the experimental detection apparatus counts atoms as
being ``ionized'' \cite{KL}.
For $n_0 = 37$, converged calculations with respect
to basis size (see below)
are obtained by using a basis of $n=30\ldots 100$.
In order to avoid reflections from our
basis end at $n=100$
we set the high $n$ part of
the wave function to $0$
after each microwave cycle.
Although the cut-off $n$ inside of the microwave
cavity is considerably higher than $n_c=91$,
we checked explicitly with computations including
up to 150 states
that this procedure is permissible, since the
chance of a highly excited state feeding back into
low-$n$ states is very small. This is consistent
with the results of early classical calculations
\cite{percival}.
The ionization probability is then calculated
according to
$P_I = 1 - P_b$, where $P_b$
is the probability to remain in the bounded states with
$n=30\ldots 90$.

The ionization probability data taken by
Koch {\it et al.} \cite{KL} are reported as a function of
the electric field amplitude
at the center of the microwave cavity. Over the
radius of the beam the amplitude of
the TM$_{020}$ mode in the microwave
cavity drops approximately 7\% \cite{KL}.
This is called radial droop. We
account for the radial droop by integrating
the atom density of the beam times the local
microwave field over the cross section of the beam.
Since we calculate the
ionization probability in steps of $1\,$V/cm,
the integration
reduces to a weighted sum.

Figure~1 shows the experimental
ionization curve together with radial droop
corrected quantum mechanical ionization curves for eight
different values of $m_0$ for $n_0 = 37$.
We see that the ionization threshold fields
increase with increasing $m_0$. This
is expected since the Stark splitting of levels
decreases with increasing $m$. Figure~1 shows that
the ionization threshold depends strongly
on $m_0$.

Taking the strong $m_0$ dependence into account is
essential for
accurately reproducing the experimental
ionization curve. Since $m$ is a
conserved quantum number and the
experiment uses an ensemble of atoms with equi-distribution
in $m$, our approach
is to find the ionization
probability for fixed $m_0$ and then average over all $m_0$.
Assuming $l_0$-independence leaves us with 37 different
calculations corresponding to
$|m_0|=0,\ldots ,36$. Each $|m_0|$
corresponds to $2(37-|m_0|)$ different
$l_0$ quantum numbers.
Since the $|m_0|$-dependence of the ionization
signal is rather smooth,
we performed calculations for
$|m_0|=0,5,10,15,20,25,30,35$ only.
Each $|m_0|\neq 0,35$
represents two ranges of $m$-values, $\{|m_0|-2,\ldots ,|m_0|+2\}$
and $\{-|m_0|-2,\ldots ,-|m_0|+2\}$, $|m_0|=0$ represents
$m=-2,\ldots ,2$ and $|m_0|=35$ represents
$m=\pm 33,\pm 34,\pm 35,\pm 36$.
The corresponding weights for the eight $m_0$ values computed
are then given by $(370-10|m_0|)/1369$ for $|m_0|\neq 0$ and
$179/1369$ for $|m_0|=0$.
Although performed in parallel on our 60-node Beowulf
cluster consisting of high-end PIII and Athlon processors
the quantum computations reported in this paper
still took several months of CPU time to complete.
This is mainly due
to the fact that we accurately included the experimental
envelope function (2) and integrated over the full number
of field cycles (about 500 per electric field value)
used in the experiments.

In addition to the quantum mechanical calculations we
performed classical molecular dynamics calculations
for the $n_0=37$ case. We used the procedures and methods
described in \cite{percival} and represented the
microcanonical distribution with 25 classical trajectories
per microwave field value.

Figure~2 shows the experimental ionization curve (broken line)
together with the results of our 3D classical (dots)
and 3D
quantum (dashed line) calculations.
As far as the main rise of the ionization signal is
concerned
both the classical and the quantum calculations
agree very well with the experimental results.
In particular, all three methods find the
onset of the main rise of the ionization signal to
occur very close to $345\,$V/cm.
The classical curve does not capture the pre-threshold
ionization structure in the experimental data.
But this is a known \cite{KL} and therefore expected
result. While the
quantum mechanical computations are in satisfactory agreement
with the experimental results as far as the main rise
of the ionization signal is concerned, the quantum
computations overestimate the ionization signal in the
pre-threshold region. We checked explicitly that while
$l_0$-independence holds to a very good approximation in
the field region of the main rise of the ionization signal,
it does not hold at all in the pre-threshold region.
This explains the discrepancy of the two ionization signals
and (i) calls for quantum calculations with explicit
$l_0$-dependence in the pre-threshold region and (ii)
points to the fact that the ionization mechanism in the
pre-threshold region is very different from the mechanism
active in the region of the main rise of the ionization signal.
Realistic quantum computations including explicit
$l_0$-dependence, however, are currently beyond our
computational means.

Discussing our results we note that the
choice of basis end is not critical.
We chose it small enough to
make the problem tractable, but large enough
to achieve convergence. This is illustrated
in Figure~3, which shows the ionization
probability for $m=0$ in bases of
$n=30 \ldots 100$ (full line)
and $n=30 ... 150$ (broken line).
Both bases yield comparable results.

We do not include any continuum states
in our quantum calculations. All
ionization occurs via excitation
of high $n$ states through the absorbing
boundary condition at $n=n_c$. In reality, of course,
all bounded states are also
directly coupled to the continuum
by the microwave field
and at first glance
neglecting these couplings seems like a bad approximation.
This is certainly true in a situation where
the main ionization channels are via direct
low-order
multiphoton transitions to the continuum. This is not
the case here where chaotic diffusion and stochastic ionization
\cite{class1,class2} are the dominant processes.
In this case ionization mainly proceeds through high-lying
Rydberg states adequately modeled
by our absorbing boundary
at $n=n_c$.
A further argument is the following.
In retrospect, assuming that the experimental results are
accurate,
this picture should be true to a good approximation.
Otherwise our ionization curves
would lie well below the experimental curves,
the difference being ``true ionization'' via direct
transitions to the continuum,
not captured by
our model.
We acknowledge, however, that
determining
the branching ratio between these
two ionization channels is an important but
challenging task.

The $n_0=37$ experiments at $9.92\,$GHz fall
into the low-frequency or adiabatic
regime (Regime-II in \cite{KL}). This
regime is characterized by structure of
quantum mechanical origin, here manifesting itself
as a prominent pre-threshold bump.
For $n_0 \geq 70$ (approximately) these
features disappear and all the
experiments are in impressive agreement
with classical 3D calculations
(Regime-III in \cite{KL}). As such, the
intermediate $n_0$ are the most
interesting to explore in the model
presented here. Still,
the $l_0$-independence
argument holds for all
$n_0$. Therefore
we expect to be able to model the
experimental data for all initial principal
quantum numbers.

In this paper
we discussed a new model for reproducing measured
microwave ionization curves of
hydrogen Rydberg atoms. We presented the
first 3D quantum mechanical calculations capable
of reproducing experimental ionization curves in
the experimentally relevant parameter regime.
In the case of $n_0=37$
we are able to accurately
reproduce the ionization threshold and
the shape of the ionization curve
following the onset of the main rise of the ionization signal.
In addition
we qualitatively capture the pre-threshold ionization
structure.

We expect that
our model works for all
of the $9.92\,$GHz microwave ionization data
so far reported in the literature \cite{KL}.
Preliminary calculations show that
equally satisfying results are obtained for
principal quantum numbers different from
$n_0=37$.

We suggest that experiments
be carried out with initial
selection of the magnetic quantum number.
We predict ionization curves from
such experiments to closely follow the curves shown in
Figure~1.
This would lend credibility to
the model presented here and allow further
studies of the details of the
ionization mechanism.

An open question is
whether narrow ``spikes''
present in the computed ionization signals \cite{CB}
are real. Indeed, much of our enthusiasm for
performing realistic 3D quantum calculations draws
from the desire to settle this question.
The question of the spikes may, of course, also be
attacked experimentally.
In current microwave ionization experiments, however,
these narrow
features are washed out by the radial droop in the
cavity field and by the presence of many
$m$ quantum numbers in the initial state.
Therefore experiments addressing the spikes
should include
$m$ selection and work with
a narrow beam in order to reduce the
radial droop of the microwave field. According to
our calculations \cite{CB} a radial droop of less than
$1\,$V/cm may be necessary.

The authors would like to thank
Peter Koch for generously providing them with
the digitized data of the experimental
ionization curves. G.N.R and V.F.H gratefully
acknowledge financial support by an NSF REU
grant. Th.C. was supported by a graduate stipend
by the Danish Research Council.
This work was supported by the NSF under Grants
No. 9900730 and 9984075.


\pagebreak

\centerline{\bf Figure Captions}

\bigskip\noindent
{\bf Fig.~1:}
Ionization probability of
microwave-driven hydrogen Rydberg atoms initially
prepared in $n_0=37$ as a function of microwave field strength.
Broken line: experimental result.
Full lines: quantum mechanical extremal Stark state calculations for
$m_0=0,5,10,15,20,25,30,35$ with radial droop correction.
The curves occur in sequence from the left-most ($m_0=0$) to
the right-most ($m_0=35$).

\bigskip\noindent
{\bf Fig.~2:}
Ionization probability of microwave-driven
hydrogen Rydberg atoms initially
prepared in $n_0=37$ as a function of microwave
field strength. Full line:
experimental result;
broken line: extremal Stark state calculation;
dots: classical calculation.
Both classical and quantum calculations are
$m$ averaged and corrected for
radial droop of the microwave field in the microwave cavity.

\bigskip\noindent
{\bf Fig.~3:}
Ionization probability of microwave-driven
hydrogen Rydberg atoms initially
prepared in $n_0=37$ and $m_0=10$ as a function
of microwave field strength.
Full line: basis $n=30\ldots 100$ (same as Figure~1).
Broken line: basis $n=30\ldots 150$.

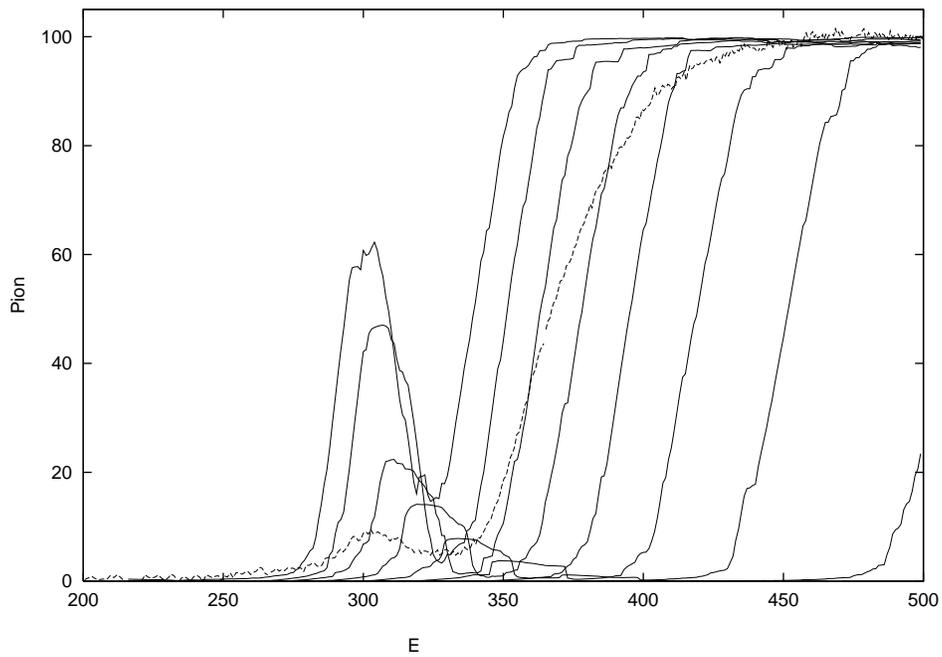
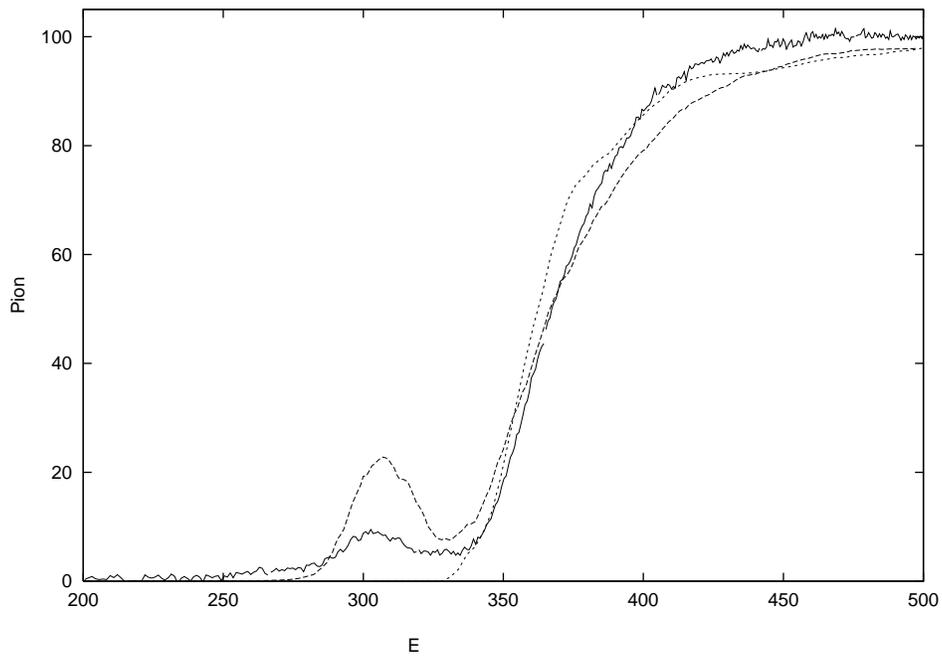
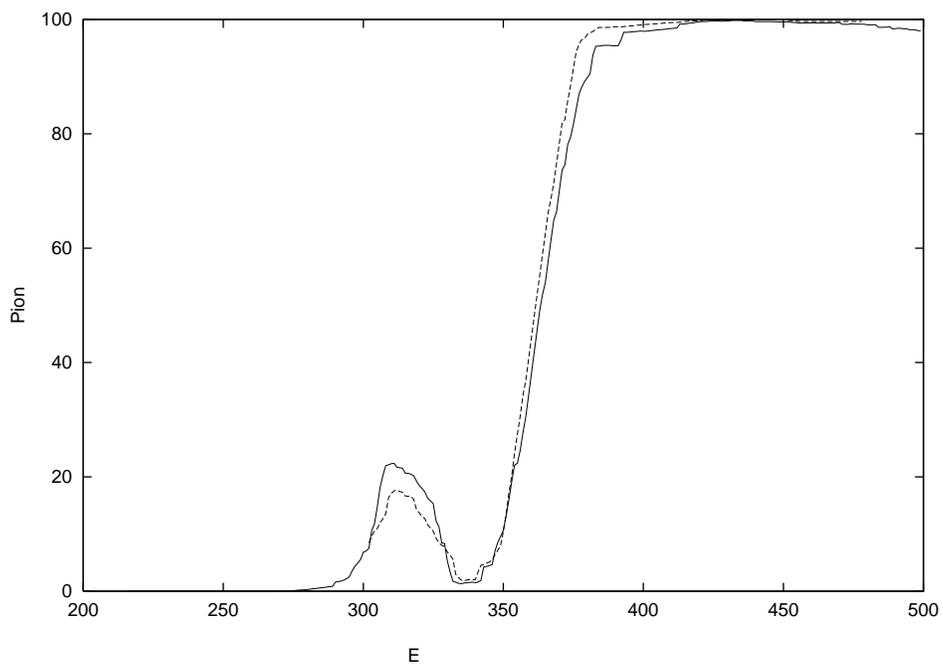
\end{document}